\documentclass[aps,prd,onecolumn,groupedaddress,showpacs,nofootinbib,amssymb
]{revtex4}
\usepackage[dvips]{graphicx}
\usepackage{amssymb}
\usepackage{amsmath}
\usepackage{graphicx,,color}
\usepackage{amsfonts}
\usepackage{bm}
\usepackage{cancel}
\usepackage{comment}

\allowdisplaybreaks[4]

\begin{document}

\tolerance=5000

\title{An analysis of the $H_{0}$ tension problem in a universe with a viscous dark fluid}

\author{
Emilio Elizalde$^{1,2}$\footnote{E-mail: elizalde@ieec.uab.es}, 
Martiros Khurshudyan$^{1,2}$\footnote{Email: khurshudyan@ice.csic.es},
Sergei D. Odintsov$^{1,3}$\footnote{E-mail: odintsov@ieec.uab.es} and
Ratbay Myrzakulov$^{4}$\footnote{Email: rmyrzakulov@gmail.com} }
\affiliation{
$^{1}$ Consejo Superior de Investigaciones Cient\'{\i}ficas, ICE/CSIC-IEEC,
Campus UAB, Carrer de Can Magrans s/n, 08193 Bellaterra (Barcelona) Spain \\
$^{2}$ International Laboratory for Theoretical Cosmology, Tomsk State University of Control Systems and Radioelectronics (TUSUR), 634050 Tomsk, Russia \\
$^{3}$ ICREA, Passeig Luis Companys, 23, 08010 Barcelona, Spain \\
$^{4}$ Eurasian International Center for Theoretical Physics and Department of General and Theoretical Physics, Eurasian National University, Nur-Sultan, 010008, Kazakhstan\\
}

\begin{abstract}

In this paper, two inhomogeneous single fluid models for the Universe, which are able to naturally solve the $H_{0}$ tension problem, are discussed. The analysis is based on a Bayesian Machine Learning approach that uses a generative process. The adopted method allows to constrain the free parameters of each model by using the model itself, only. The observable is taken to be the Hubble parameter, obtained from the generative process. Using the full advantages of our method, the models are constrained for two redshift ranges. Namely, first this is done  with mock $H(z)$ data over $z\in [0,2.5]$, thus covering known $H(z)$ observational data, which are most helpful to validate the fit results. Then, aiming to extend to redshift ranges to be covered by the most recent ongoing and future planned missions, the models  are constrained for the range $z\in[0,5]$, too. Full validation of the results for this extended redshift range will have to wait for the near future, when higher redshift $H(z)$ data become available. This makes our models fully falsifiable. Finally, our  second model here is able to explain the BOSS reported value for $H(z)$ at $z=2.34$.
\end{abstract}

\pacs {}

\maketitle

\section{Introduction}\label{sec:INT}

The discovery of the accelerated expansion of our Universe is one of the greatest events of our epoch~\cite{Riess}~-~\cite{Hinshaw}. It requires the existence of some kind of dark energy, to be added to our working models of the Universe. However, over twenty years have elapsed and the nature of dark energy is still a mystery; although we hope that this could changed at any point, due to the high quality observational data expected from future collaborations and the corresponding space missions~\cite{Inserra}~-~\cite{Sartoris}~(to mention a few). On the other hand, available observational data already provide some hints, partially shedding  light on this dark energy problem. Nowadays, from the observational data alone (in a model-independent way), it is possible to decide what are the directions to look for, and the possible candidates for dark energy to be introduced, in order to solve the acceleratedly expanding Universe problems~(for dark energy models with fluids, see~\cite{Bamba:2012cp}~-~\cite{Arman} and references therein). The most interesting aspect in all these tools to be mentioned here is, that different approaches give almost the same result, and it is practically impossible to take a final decision, discriminating among them. There are different reasons why the dark energy problem cannot be  solved, one among the main being the presently existing tension between different observational data-sets. This shades reasonable doubts, leading to believe that the models in hand have real problems, and that these problems can only be fixed with new observational data covering extended redshift ranges. Indeed, it seems correct to say that only the data from higher-redshift observations can solve the problems at hand, because we already have qualitatively good data covering the low-redshift behavior of our Universe. The mentioned tension problem seems to come from these low-redshift data and a significant part of the research seems to have neglected this fact and, at the same time, assumed that the error distribution is also Gaussian. Removing one or more of these assumptions may lead to different outcomes. However, the existing tools commonly used in the model analysis are too insensitive or either cannot be used, when one of the assumptions is changed or removed. On the other hand, at this moment it is even impossible to predict if the new data will resolve the tension between the data-sets or if the existing tension will be masked, but more complicated tasks to be solved will appear. Anyway, let us for a while forget about the tension issue and briefly summarize what we know about dark energy and the accelerated expansion problems.

We know that if General Relativity is the correct theory of gravity to describe the background dynamics of our late time Universe, then we need some energy source able to accelerate its expansion. On the other hand, we know that it should be an energy source behaving in such a way that it will not destroy the observable Universe full of well-studied, different objects. Moreover, if we take the cosmological constant model to be the dark energy model, then it soon becomes clear that we must be ready to perform modifications, because it suffers of very serious problems. Soon after the discovery of the cosmological constant problems it was natural to assume that the dark energy has a dynamical nature, and  various interacting dynamical dark energy models where  introduced and developed. Nowadays, intensive work in this direction is taking place, and there is the hope that the new problems may be eventually solved by some interacting dynamical dark energy model\footnote{We refer the readers to \cite{Bamba:2012cp}~-~\cite{Arman}  and references therein for further detailed discussion about fluid dark energy with related problems and its possible resolution.}. However, we should mention another interesting late time Universe related problem can be an efficient way of revealing the nature of dark energy. In the recent literature, there is an intensive discussion about a recent problem, known as the $H_{0}$ tension issue. It arose from the Planck and Hubble Space Telescope result reports, when the community noticed that, on top of the other problems of modern cosmology, there is also a tension between the caculated values of $H_{0}$ obtained from the two observational data sets. The problem is due to a reported huge difference between the observed $H_{0}$ values coming from the two mentioned projects. This could have two reasons, indicating either new physics, or a problem with the measurements. For the moment, it is very hard to estimate what to expect; however, one thing seems clear --if these results indicate new physics, then we have a very good starting point to clarify, constrain and select good extensions and modifications of General Relativity. A recent discussion on the problem  drew up the borders separating good models from  bad ones. However, it is too early to reach a final conclusion, since the measurement-related issues and their strong impact on the $H_{0}$ tension problem have not been excluded completely, as of now. It has been shown that some of the interacting dynamical models existing on the market can be used to solve the issue. On the other hand, a modification of General Relativity can also propose solutions~\cite{H0Start}~-~\cite{H0End}~(to mention some references). In this paper we discuss another approach to address this issue.

Specifically, the main goal of our study is to address the $H_{0}$ tension problem as a good chance to construct new cosmological models with viscous/inhomogeneous fluids~\cite{INStart}~-~\cite{INEnd}. Indeed, nowadays, in the scientific literature there is a huge amount of works indicating the ways how the $H_{0}$ tension problem can be solved, or at least alleviated. However, up to our knowledge, there is no work connecting the $H_{0}$ tension problem solution with inhomogeneous single-fluid Universe models. It should be mentioned that a previous study shows that cosmic fluid viscosity/inhomogeneity could be very useful not only for modeling the accelerated expanding Universe, but also for modeling cosmic inflation. Therefore, it is natural to expect that viscosity and inhomogeneity of the cosmic fluid can be useful in this case too. However, an early study with existing models has shown that we needed to construct a new one, since the old models could not solve the $H_{0}$ tension issue. In this paper we will prove that, by using Bayesian Learning and probabilistic programming tools, we are able to construct new models of an inhomogeneous single fluid Universe were the $H_{0}$ tension problem disappears. In particular, we will present two new models able to explain both the accelerated expansion of the late time Universe and the transition to this phase. As we have mentioned, our study is based on a Bayesian Machine Learning approach, which actually does not require real observational data to be used for the analysis. The method employs a model based generative process, allowing to constrain the free parameters of the model. In our case, the observable used in the Bayesian Learning process is taken to be the Hubble parameter and, using the advantages of the method, we manage to constrain the models for two redshift ranges, each. Namely, first we constrain it with generative-process based $H(z)$ data over $z\in [0,2.5]$, thus covering well-known $H(z)$ observations. Those will be helpful to validate our fit results. On the other hand, taking into account the redshift ranges to be covered by ongoing and future planned missions, we constrain the models for $z\in[0,5]$. The validation of our results for the second redshift range will be done in the future with to-be-measured higher redshift $H(z)$ data.  

Before ending this section, we recall the standard notation of FLRW cosmology~(with $8 \pi G = c = 1$). In particular,  the metric in this case has the  form
\begin{equation}
ds^{2} = -dt^{2} + a(t)^{2} \sum_{i =1}^{3} (dx^{i})^{2},
\end{equation}
and  
\begin{equation}\label{eq:F1}
H^{2} = \frac{1}{3}\rho.
\end{equation}
Moreover, the energy conservation law can be expressed as
\begin{equation}\label{eq:EC}
\dot{\rho} + 3 H (\rho + P) = 0.
\end{equation}
On the other hand, the last equation is equivalent to the following 
\begin{equation}\label{eq:F2}
\dot{H} = -\frac{1}{2} (\rho + P).
\end{equation} 
In the above three equations $H$, $P$ and $\rho$ are the Hubble parameter, the pressure and the energy density of the inhomogeneous fluid, respectively. 

The paper is organized as follows. The philosophy behind the  method here used is discussed in Sect.~\ref{sec:BIBBL}. In Sect.~\ref{sec:BEM}, we present the models and discuss the constraints obtained, followed by an analysis of the consequences, interesting for the $H_{0}$ tension and for the accelerated expanding Universe issues. The model based generative process employed in the Bayesian Learning approach is built from Eqs.~(\ref{eq:F1}) and~(\ref{eq:EC}), by assuming new forms of $P = P(\rho)$ to describe the inhomogeneous single fluid Universe. The model based generative process and the analysis leading to our final results has been performed by using PyMC3\footnote{https://docs.pymc.io}. The final conclusions of our analysis are given  in Sect.~\ref{sec:conc}. Additionally, there is an Appendix  with some details of the method.

\section{Basic ideas behind Bayesian Learning}\label{sec:BIBBL}

An important question nowadays is to understand how efficiently one can deal with the unprecedented scale and resolution data expected from future  missions and collaborations. However, the extended scale and high resolution of the expected data should be first defined, in order to be sure that the new algorithm may be sufficient to do the task. But since this is a generic computing and data science problem, we leave any future discussion on this point to the specialists on this field, and we will just address the issues which, according to some estimations, fiel into the cosmology and astrophysics domains. It is well known that  mathematical models are very useful in understanding nature and that conveniently adjusting their free parameters is very important and should be done in a very efficient way. In particular, it should be done also for cases when we have only a few observational data-points or the quality of the observations is not good, causing huge errors. In other words, the mathematical model at hand, first of all, should be made consistent with observations by adjusting the free parameters and this should be done efficiently. At the same time, often this must be done in  practice for observational data-sets without relying too much on the quality of the data. Several years ago, this was really a very challenging task for computer science and, consequently, imposed severe restrictions on different fields as, among them, cosmology and astrophysics~\cite{MLStart}~-~
\cite{MLEnd}. However, nowadays thanks to significant developments in different scientific fields the mentioned difficulties have been alleviated, to some extent. We already mentioned that the adjustment of model-free parameters is an important step in any study. On the other hand, the comprehensive analysis of the model includes also the estimation of the errors of the same parameters, still providing an observational data consistent behavior shown by a certain mathematical model. Moreover, if we have a well-studied model, it can be useful and extremely informative for designing next-generation experiments and observations. The increasing number of research in this direction seems to indicate that machine learning algorithms can be indeed used efficiently to perform the model fitting and overcome observational data issues, as the ones mentioned above. Before describing the basic ideas behind the approach used in this paper to study the cosmological models, let us briefly present typical machine learning procedures used to study cosmology and astrophysics, as discussed in the recent literature. In this discussion, we will not describe any specific procedure used to extract observational data and will not introduce a specific neural network used in the model analysis. We wish to keep the level as simple as possible. We would just like to mention that, in general, we follow these  three steps:
\begin{itemize}

\item It is obvious that, first of all, we should define the model. Usually in this step we get or define a set of equations/rules controlling the model behavior. In other words, the equations link the model parameters together, according to some rule allowing to study the behavior of the model at different regimes.

\item Usually after this, we need to choose a set of data to find the free parameters. In physics, nearly in all cases the data would be obtained either from some experiment or observation, i.e. we will use data related to some real physical processes. Actually, another interesting situation to be mentioned is the case when the data may be simulated; but, at this moment, we cannot validate simulated data because the related experiment/observation setup is not operating or is still under construction. This is still intensively discussed in the recent literature, about future missions .   

\item Finally, when the first two steps have been done, we run a learning algorithm. In other words, we use data to determine the values for the unknown model parameters. Now what does it mean that we run a learning algorithm? This is the most interesting step, to be clarified, because we present the data in terms of input and output pairs and then run some optimization algorithms to get a final set of weights; i.e., we train the network allowing to determine free parameters.
\end{itemize}

Now, having defined the basic steps behind machine learning studies, intensively used in the literature, let us see what is behind the approach used in this paper.  Our approach is just known as Bayesian Learning and has been implemented using PyMC3 probabilistic programming, a python-based framework, which in practice has proven to be very fast, useful, and able to be easily integrated with other python-based frameworks~\cite{PyMC3}. Indeed,  a lot of effort has been put on the PyMC3 project before, so that we can mostly concentrate our attention here on the physics behind the problem under study. In order to better understand what is behind Bayesian Learning, we need to modify and adapt the above mentioned $3$ steps to this specific situation and, as a consequence, we now have:

\begin{itemize}

\item We need to define the model, to be used to provide a so-called generative process. It is needed to generate the data and this means that, by defining the model, we will define a sequence of steps describing how the data was created. It is clear that the generative process includes the unknown model parameters and uses explicit values of these parameters. Therefore, the crucial aspect here is the incorporation of our prior beliefs to the unknown parameters. Eventually, we need only prior information to get the posterior and to constrain the parameters, since we use learning algorithms we do not need, anymore, to evaluate the likelihood, as in the case of $\chi^{2}$ analysis, up to now widely used to constrain cosmological models.

\item The crucial step happens here, at this stage, when we envisage the data to be the data obtained from the generative process.

\item Eventually, after running the learning algorithm, we update our belief about the parameters and get a brand new distribution over these parameters. 

\end{itemize}

It should be mentioned that the Likelihood-Free inference methods allow us to perform Bayesian Inference using forward simulations only, with no reference to a likelihood function. This is particularly appealing for cosmological data analysis problems, where complex physical processes and instrumental effects can often be simulated; but incorporating them into a likelihood function and solving the inverse inference problem is much harder. Likelihood-Free methods generically require large datasets to be compressed down to a small number of summary statistics, in order to be scalable. On the other hand, Approximate Bayesian Computation (abc) approaches to Likelihood-Free inference draw parameters from the prior, and forward simulate mock data, accepting points whenever the simulated data fall inside some small-ball around the observed data. Together, massive data compression and density estimation Likelihood-Free Inference provides a framework for performing scalable Likelihood-Free inference from large
cosmological datasets, even when forward simulations are computationally expensive. This opens the door to a new paradigm for principled, simulation-based Bayesian inference in cosmology and astrophysics, among other relevant research fields.

PyMC3 is one of the python-based frameworks providing a comprehensive set of pre-defined statistical distributions that can be used as model building blocks~\cite{PyMC3}. It uses Theano\footnote{http://deeplearning.net/software/theano}, that is, a deep learning python-based library, to construct probability distributions and then access  the gradient, in order to implement cutting edge inference algorithms. Of course there are other useful frameworks, as well; but we shall here concentrated our attention on PyMC3, because it allows to write down models using an intuitive syntax to describe a data generating process. It allows to fit the model using gradient-based MCMC algorithms for fast approximate inference, or to use Gaussian processes in order to build Bayesian nonparametric models. Indeed PyMC3 gives all necessary tools for the analysis, allowing to concentrate our attention  on the real problem, only.

The purpose of this section was to provide the general philosophy behind the method used in this paper to study the $H_{0}$ tension problem, within the scheme of an inhomogeneous single fluid Universe. We have omitted any specific discussion on the mathematical framework behind deep learning algorithms and Bayesian Learning, since these  can all be found in PyMC3 standard tutorials, which are endowed with numerous excellent examples demonstrating how the above  ideas can actually be implemented within a practical problem.

\section{Two single-fluid models and their corresponding results}\label{sec:BEM}

In this section we present our models of an inhomogeneous, single-fluid Universe, and discuss the results obtained. In order to simplify the presentation, we have organized it into two subsections. To summarize the fit results and make them easily readable, for both models we organize corresponding tables, Table~\ref{tab:Table1}, for $z\in [0,2.5]$ and for $z \in [0,5]$, respectively. 

\begin{table}
  \centering
    \begin{tabular}{ | c | c | c | c | c | c |  p{2cm} |}
    \hline
    
 Model 1 & $H_{0}$ & $\omega_{0}$ & $\omega_{1}$ & $A$ & $n$ \\
      \hline
      
 when $z\in[0,2.5]$ & $73.395 \pm 0.1$ km/s/Mpc & $0.608 \pm 0.007$ & $-1.637 \pm 0.01$ &  $-1.01 \pm 0.01$ & $ 0.55 \pm 0.01$\\
          \hline
when $z\in[0,5]$  & $73.393 \pm 0.1$ km/s/Mpc& $0.614 \pm 0.0051$ & $-1.549 \pm 0.01$ &  $-0.95 \pm 0.01$ & $ 0.49 \pm 0.01$\\

           \hline
 
 \multicolumn{6}{c}{} \\ \hline
 
 Model 2 & $H_{0}$ & $\omega_{0}$ & $\omega_{1}$ & $A$ & $n$ \\
       \hline

when $z\in[0,2.5]$ & $73.52 \pm 0.152$ km/s/Mpc & $0.55 \pm 0.0055$ & $-1.605 \pm 0.0087$ & $-0.98 \pm 0.0098$ & $0.75 \pm 0.01$  \\
         \hline
 when $z\in[0,5]$  & $73.6 \pm 0.146$ km/s/Mpc & $0.511 \pm 0.0044$ & $-1.501\pm 0.0088$ & $-0.98 \pm 0.01$ & $ 0.75 \pm 0.01$  \\   
     \hline
    
    \end{tabular}
\caption{Best fit values and $1\sigma$ errors estimated for Model 1, Eq.~(\ref{eq:Model1}), and for Model 2, Eq.~(\ref{eq:Model2}), for $z \in [0,2.5]$ and $z \in [0,5]$, respectively. The results has been obtained from a Bayesian Learning approach, where the generative based process has been organized using Eq.~(\ref{eq:F1}) and Eq.~(\ref{eq:EC}), assuming new forms of $P = P(\rho)$ to describe the inhomogeneous single fluid Universe.}
  \label{tab:Table1}
\end{table} 

\subsection{Model 1}

\begin{figure}[h!]
 \begin{center}$
 \begin{array}{cccc}
\includegraphics[width=120 mm]{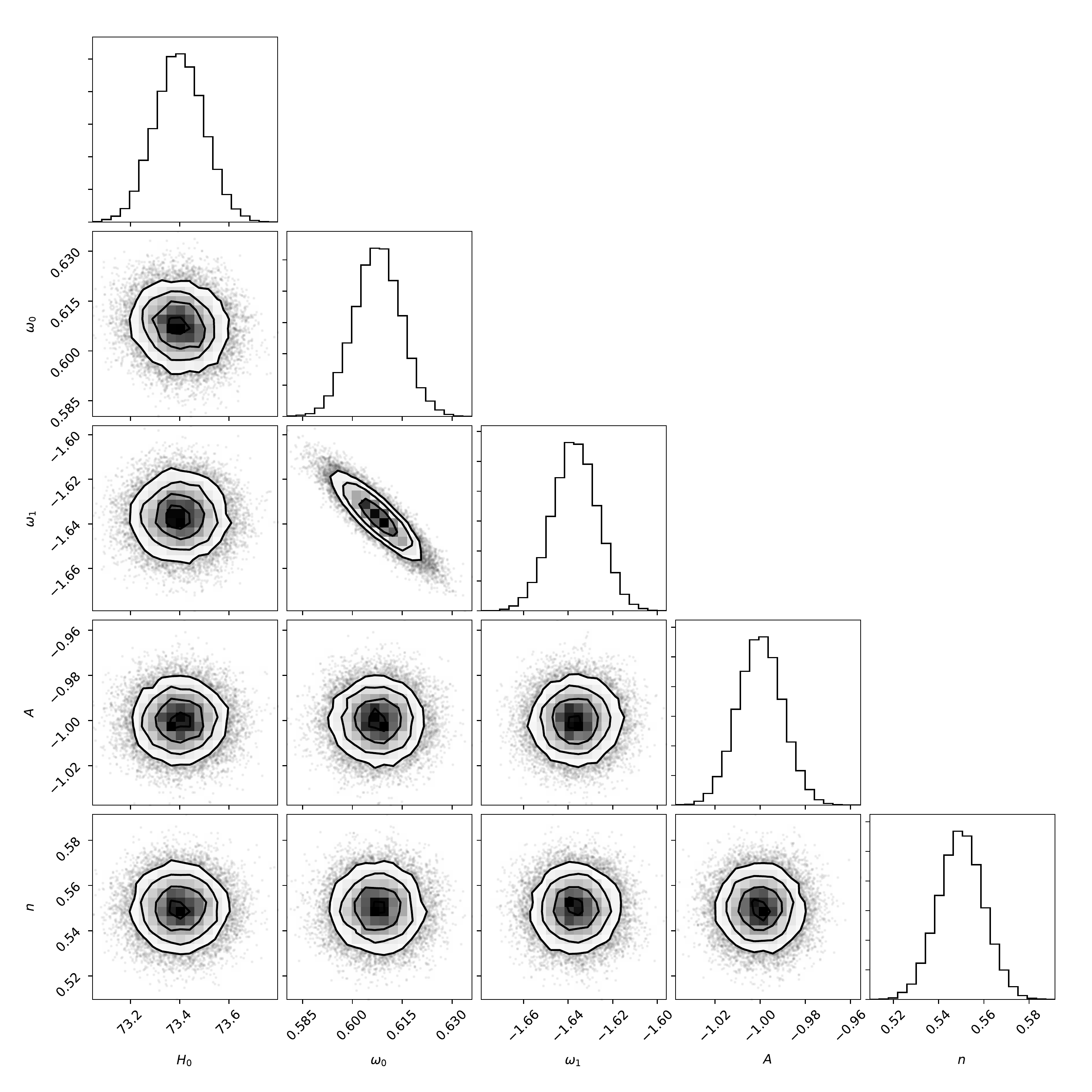}
 \end{array}$
 \end{center}
\caption{Contour map of the model given by Eq.~(\ref{eq:Model1}) for $z \in [0,2.5]$. The best fit values of the model parameters have been found to be $H_{0} = 73.395 \pm 0.1$ km/s/Mpc, $ \omega_{0} = 0.608 \pm 0.007$, $\omega_{1} = -1.637 \pm 0.01$, $A = -1.01 \pm 0.01$, and $n = 0.55 \pm 0.01$, when a Bayesian Learning approach based on the generative process has been applied.}
 \label{fig:Fig0_1_a}
\end{figure}

The first example of the inhomogeneous single fluid Universe we studied would be described by the following EoS 

\begin{equation}\label{eq:Model1}
P = \left ( \omega_{0} + \frac{\omega_{1}}{1+z} \right ) \rho - A H^{n},
\end{equation}
where $\omega_{0}$, $\omega_{1}$ and $n$ are the free parameters of the model, while $P$ and $\rho$ are the pressure and the energy density of the inhomogeneous equation of state of the Universe, respectively. In order to organize the generative process based Bayesian Learning, we need to combine Eqs.~(\ref{eq:F1}) and~(\ref{eq:EC}) with Eq.~(\ref{eq:Model1}), and take into account that $\dot{\rho} = - (1+z) H \rho^{\prime}$. In particular, from Eq.~(\ref{eq:F1}) we have that $\rho = 3 H^{2}$, therefore $\dot{\rho} = 6 H \dot{H}$, which together with  Eq.~(\ref{eq:EC}) and after some algebra will eventually  yield  the following differential equation:
\begin{equation}\label{eq:HP1}
H^{\prime} = \frac{3 H^2 (\omega_{0}+ \omega_{1} + \omega_{0} z + z + 1)- A (z+1) H^{n}}{2 H (z+1)^2},
\end{equation}
where the prime denotes the redshift derivative of the function. Namely the last equation describes the observable associated with our model and it namely has been used in the generative process. The performed study puts the constraints on the model free parameters and yields the results describe below. In particular, first of all, we observe that the Bayesian Learning puts very tight constraints on the parameters. This can be seen from Table~\ref{tab:Table1} and from the ensuing presentation 
\begin{itemize}

\item The best fit values of the model parameters are $H_{0} = 73.395 \pm 0.1$ km/s/Mpc, $ \omega_{0} = 0.608 \pm 0.007$, $\omega_{1} = -1.637 \pm 0.01$, $A = -1.01 \pm 0.01$ and $n = 0.55 \pm 0.01$ when $z \in [0,2.5]$. The contour map, when $z \in [0, 2.5]$, is given in Fig.~(\ref{fig:Fig0_1_a}). Moreover, using the best fit values we have found that, at $z = 0$, we would have $P/\rho = -1.028$ and $q = -1.042$.

\item  The best fit values of the model parameters will most likely be $H_{0} = 73.393 \pm 0.1$ km/s/Mpc, $ \omega_{0} = 0.614 \pm 0.0051$, $\omega_{1} = -1.549 \pm 0.01$, $A = -0.95 \pm 0.01$, and $n = 0.49 \pm 0.01$, when $z \in [0,5]$. The contour map, when $z \in [0; 5]$, is represented in Fig.~(\ref{fig:Fig0_1_b}). Using the best fit values, we have found that, for $z = 0$, the most likely values are: $P/\rho = -0.935$ and $q = -0.902$.

\end{itemize}

\begin{figure}[t!]
 \begin{center}$
 \begin{array}{cccc}
\includegraphics[width=80 mm]{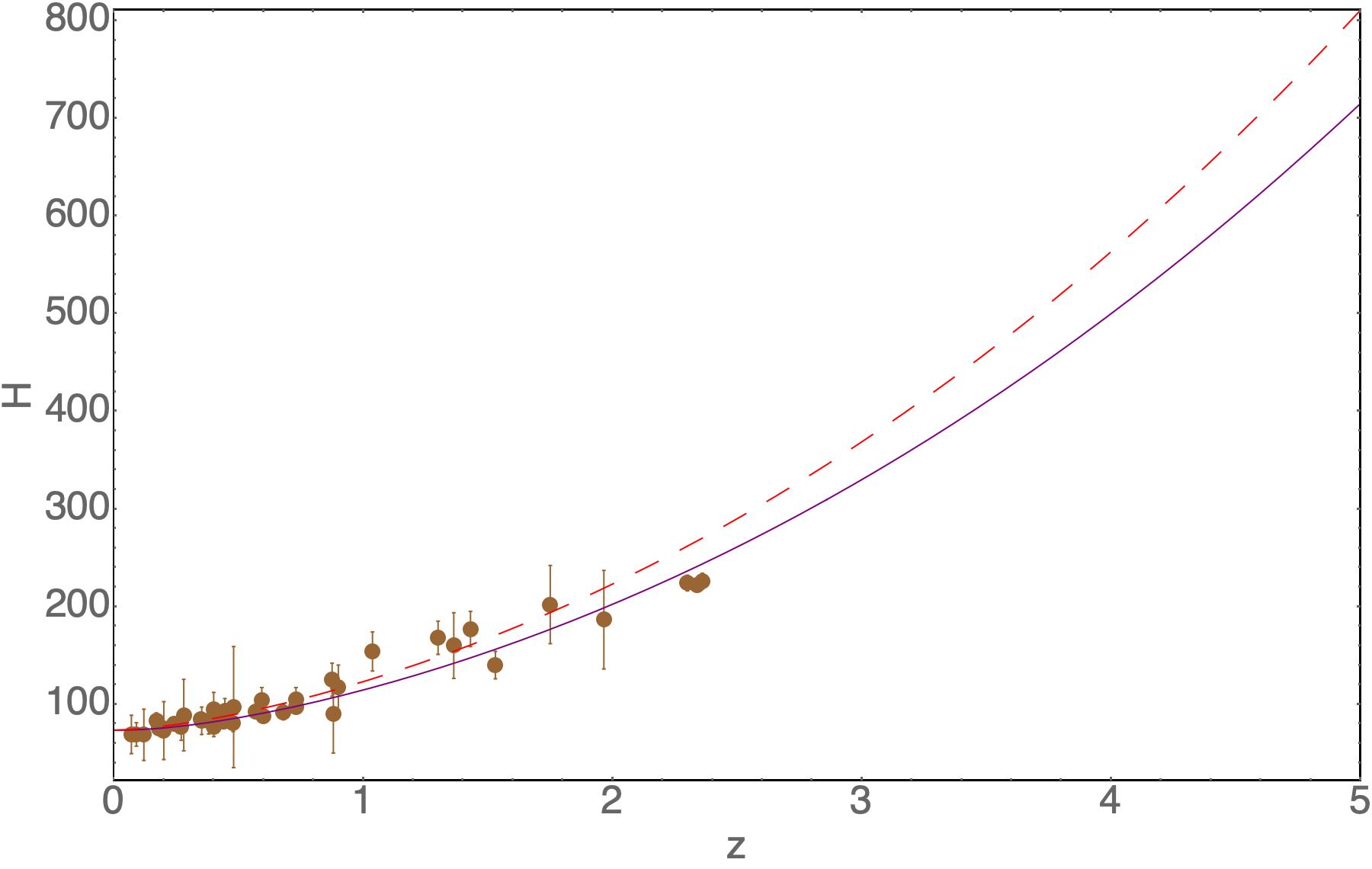}&&
\includegraphics[width=80 mm]{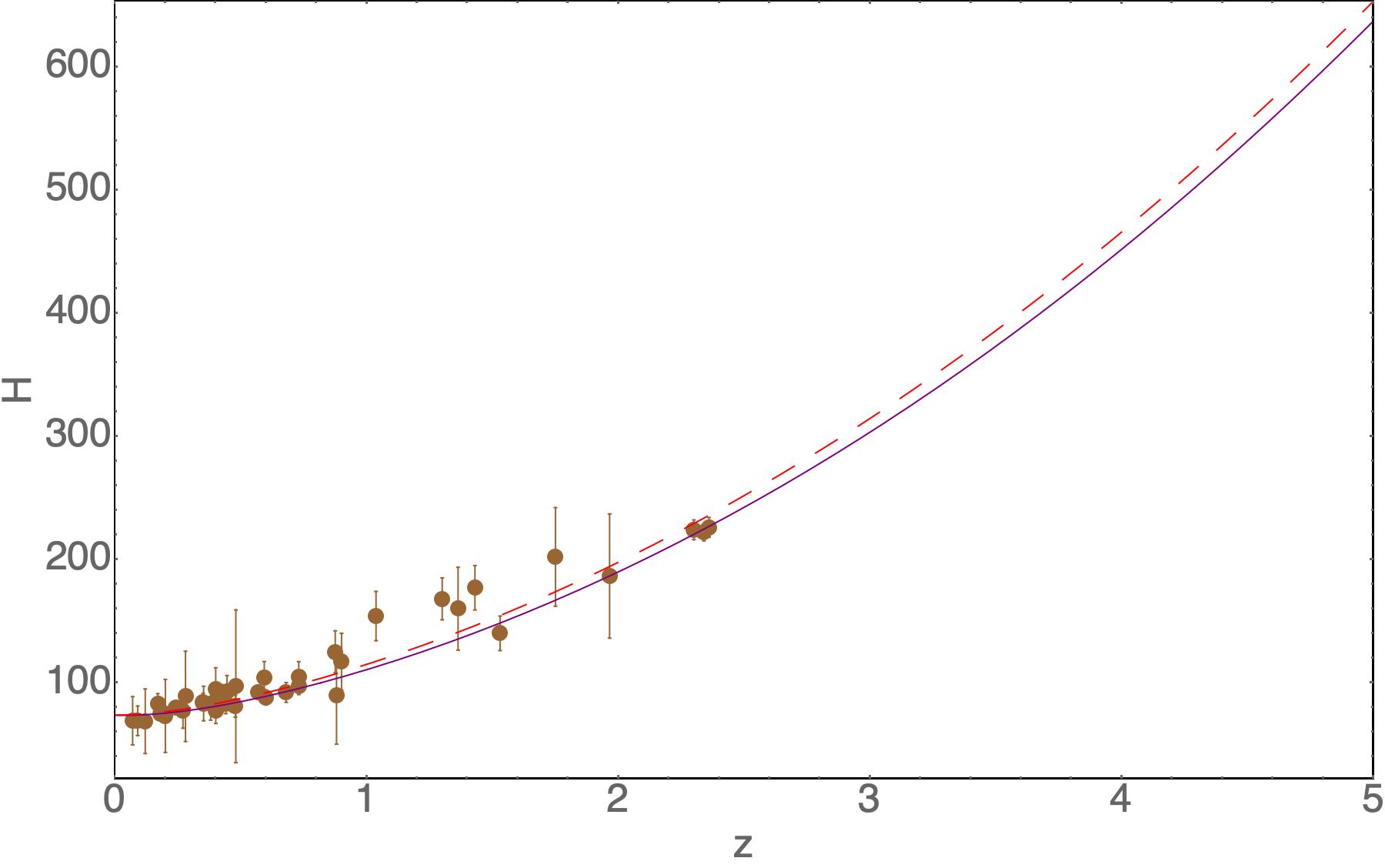}\\
 \end{array}$
 \end{center}
\caption{Graphical behavior of the Hubble parameter in comparison with known $H(z)$ data. The purple curve is a plot of the Hubble parameter for the best fit values of the model parameters, when $z \in [0,2.5]$, while the dashed red curve has been chosen for the case when $z \in [0,5]$. The red dots represent the known observational $H(z)$ data, and it is the same as in Ref.~\cite{ModifStart_3}. The left-hand side represents the comparison for the model given by Eq~(\ref{eq:Model1}), while the right-hand side plot represents the comparison with the model given by Eq~(\ref{eq:Model1}). In both cases, only the best fit values for the model free parameters obtained by the Bayesian Learning approach have been used.}
 \label{fig:Fig0_2}
\end{figure}

\begin{figure}[h!]
 \begin{center}$
 \begin{array}{cccc}
\includegraphics[width=80 mm]{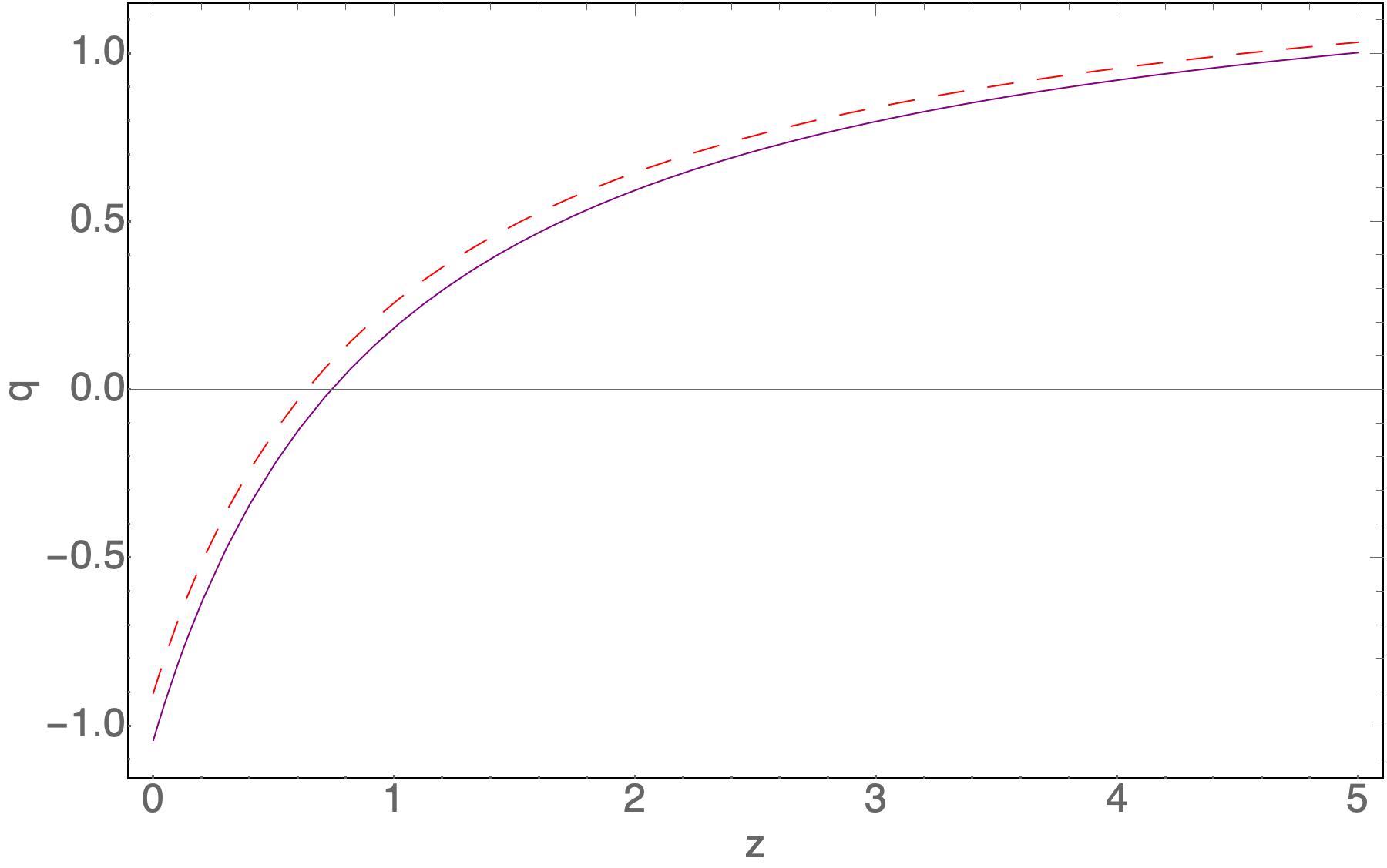}
\includegraphics[width=80 mm]{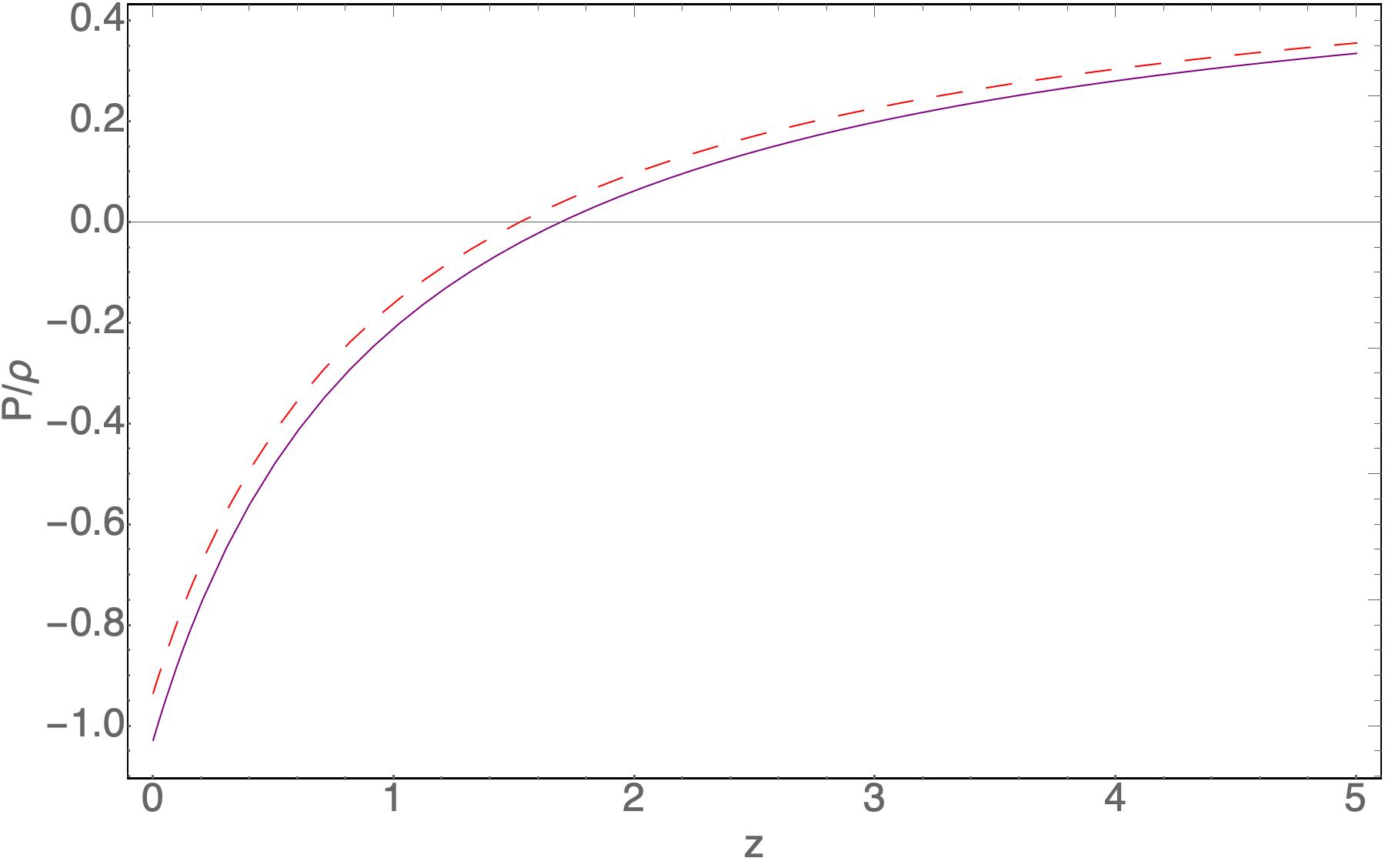}\\
\includegraphics[width=80 mm]{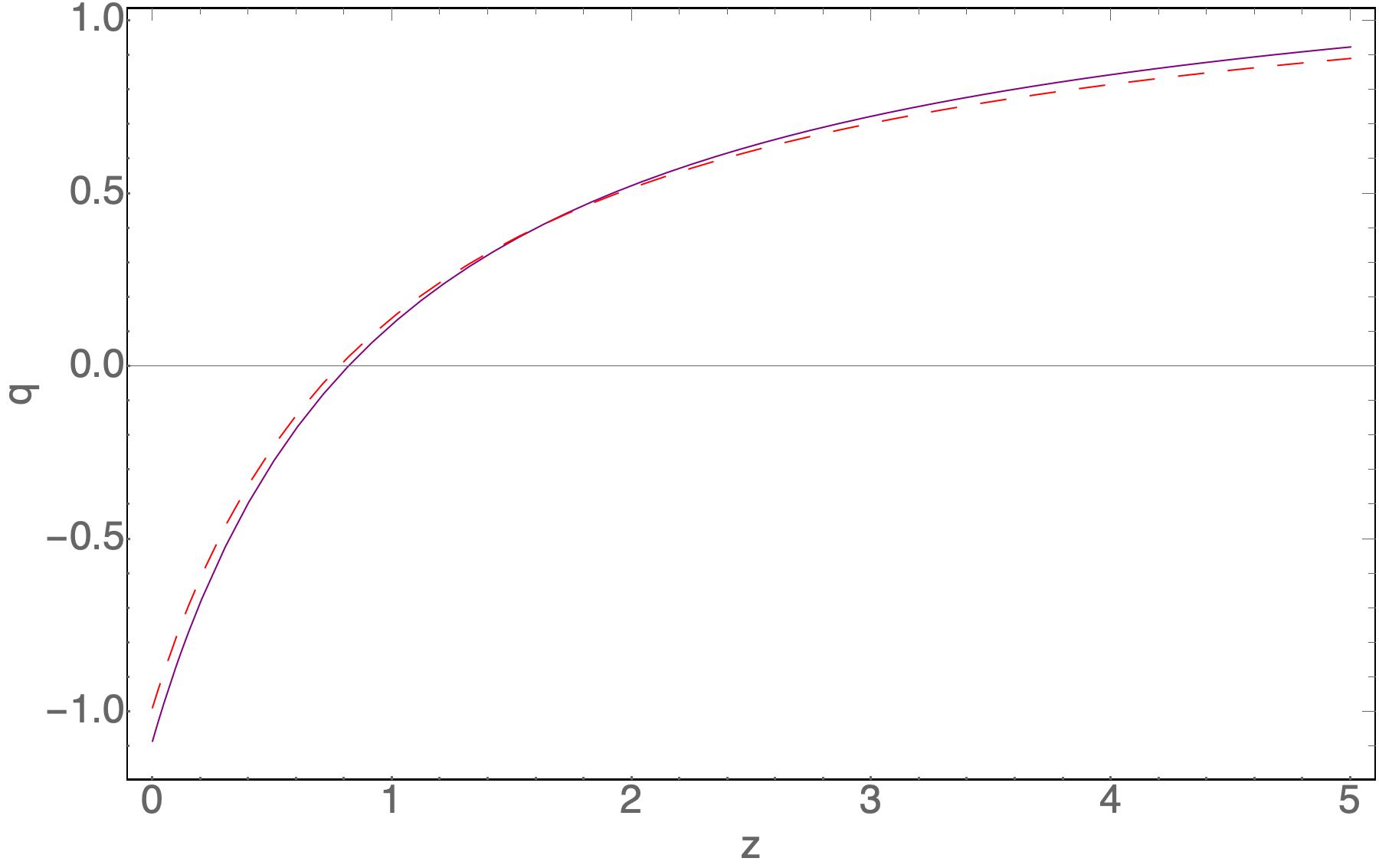}
\includegraphics[width=80 mm]{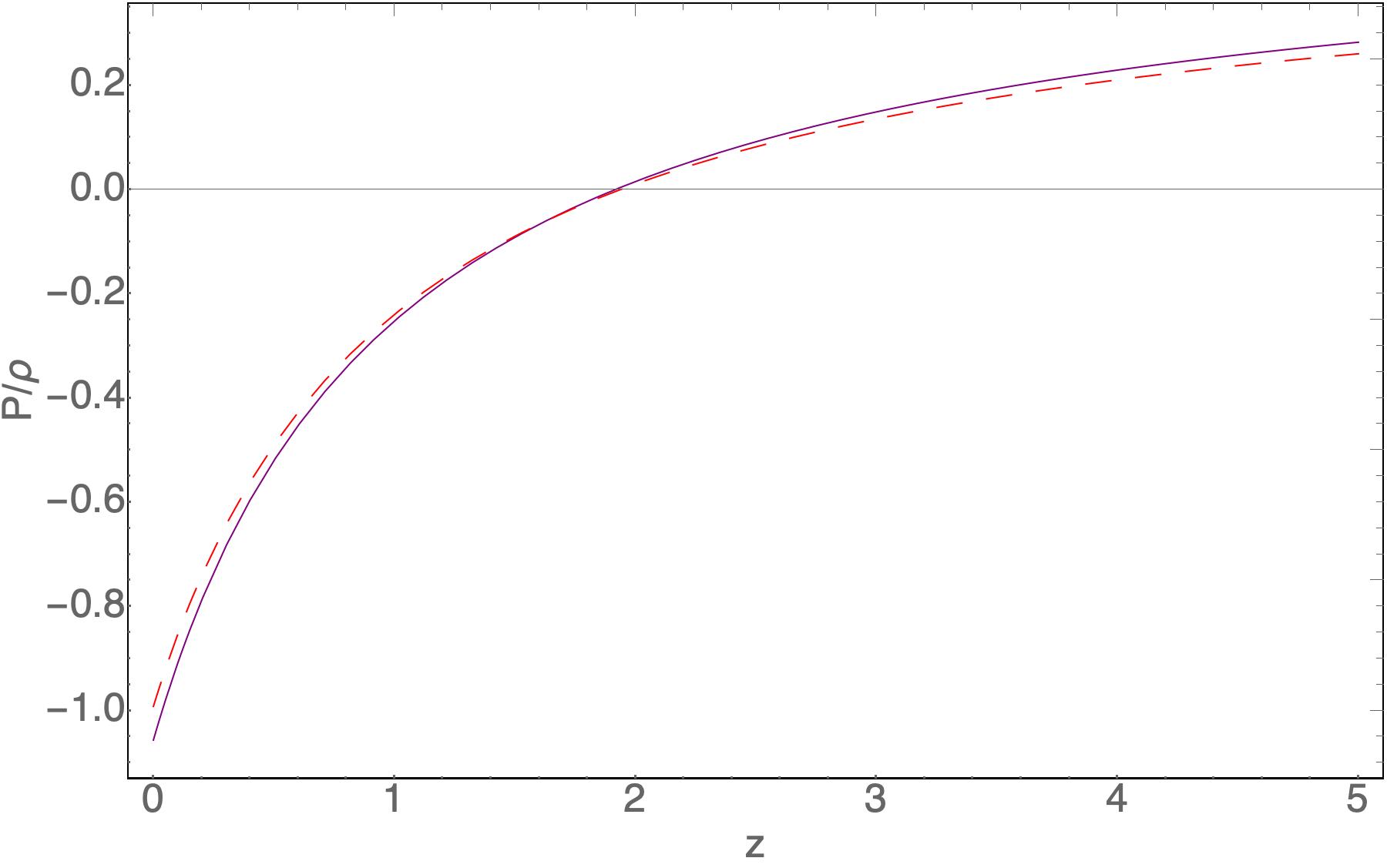}
 \end{array}$
 \end{center}
\caption{Graphical behavior of the deceleration parameter $q(z)$ and $\omega(z) = P/\rho$ for the model given by Eq.~(\ref{eq:Model1}) is given by the top panel, while the bottom panel represents the graphical behavior of the same parameters when the model is given by Eq.~(\ref{eq:Model1}). In both cases, only the best fit values of the model parameters have been taken into account.  Namely, the purple curve represents the graphical behavior of the deceleration and $\omega(z)$ parameters for the best fit values of the model parameters, when $z\in[0,2.5]$. On the other hand, the dashed red curve represents the graphical behavior of the same parameters for the best fit values of the model parameters, when $z\in[0, 5]$ has been considered. The best fit values used in these plots, when the model is given by Eq.~(\ref{eq:Model1}), are:  $H_{0} = 73.395$ km/s/Mpc, $ \omega_{0} = 0.608$, $\omega_{1} = -1.637$, $A = -1.01$ and $n = 0.55$  for $z \in [0,2.5]$, while $H_{0} = 73.393$ km/s/Mpc, $ \omega_{0} = 0.614$, $\omega_{1} = -1.549$, $A = -0.95$, and $n = 0.49$,  for $z \in [0,5]$. The best fit values used in the plots, when the model is given by Eq.~(\ref{eq:Model2}), are: $H_{0} = 73.52$ km/s/Mpc, $ \omega_{0} = 0.55$, $\omega_{1} = -1.605$, $A = -0.98$, and $n = 0.75$, for $z \in [0,2.5]$, while $H_{0} = 73.6 \pm 0.146$ km/s/Mpc, $ \omega_{0} = 0.511 \pm 0.0044$, $\omega_{1} = -1.501\pm 0.0088$, $A = -0.98 \pm 0.01$ and $n = 0.75 \pm 0.01$, for $z \in [0,5]$. }
 \label{fig:Fig0_3}
\end{figure}

Moreover, and first of all, in order to validate our results obtained from the Bayesian Learning approach, we compare the behavior of the Hubble parameter $H(z)$ obtained from Eq.~(\ref{eq:HP1}) with the available observational $H(z)$ data. The result of this comparison for the best fit values of the model parameters is given in Fig.~(\ref{fig:Fig0_2})~(left-hand side plot). From this, we see that our model can indeed explain the low redshift $H(z)$ observations very well, but most likely, some tension could arise with high redshift observations. It is only reliable for the data corresponding to $z \in (2,2.4]$. The graphical behavior of the deceleration parameter and the equation of state parameter of the fluid, Eq.~(\ref{eq:Model1}), can be found in Fig.~(\ref{fig:Fig0_3})~(top panel). In all cases, only the best fit values of the model parameters have been taken into account.  Moreover, the purple curve has been taken to represent the case when $z\in[0,2.5]$, while the dashed red curve represents the case when $z\in[0, 5]$. It should be mentioned that the model can explain the accelerated expansion and the transition to this phase. Moreover, we can see that our fluid model, Eq.~(\ref{eq:Model1}), during the evolution, will naturally evolve from a fluid with $P  > 0$ to $P < 0$, i.e. the dynamical nature of the cosmic fluid is caused to the emergence of dark energy, which is responsible for the late time accelerated expansion of the Universe. 

Before discussing how the future measurements of the expansion rate from $z \in [0,5]$ will affect the fit results, let us see what is the conclusion about the $H_{0}$ tension problem. Indeed, we can observe that the model can efficiently solve this problem, and the estimations of the $P/\rho$ EoS and of the deceleration parameter $q$ at $z = 0$, indicate that the suggested model is a viable cosmological model. On the other hand, the constraints obtained when $z\in [0,5]$ indicate that the $H_{0}$ value will, most likely, not be affected; however, the mean of the other parameters may indeed be affected, and significantly. The last may crucially affect the values of $P/\rho$ and $q$, at $z=0$.  Subsequently, this will also affect the transition redshift.  

To end this subsection, we would like to mention, again, that we have considered an inhomogeneous single-fluid model for the Universe, where the $H_{0}$ tension problem can indeed be solved. However, some tension can still raise with available high-redshift $H(z)$ data. Anyway, in the next subsection we present another viable cosmological model that can also solve this $H_{0}$ tension, so that finally there is no tension with high-redshift values. Moreover, most likely, even with the new, high-redshift expansion rate observations, the model may still remain viable.

\begin{figure}[h!]
 \begin{center}$
 \begin{array}{cccc}
\includegraphics[width=120 mm]{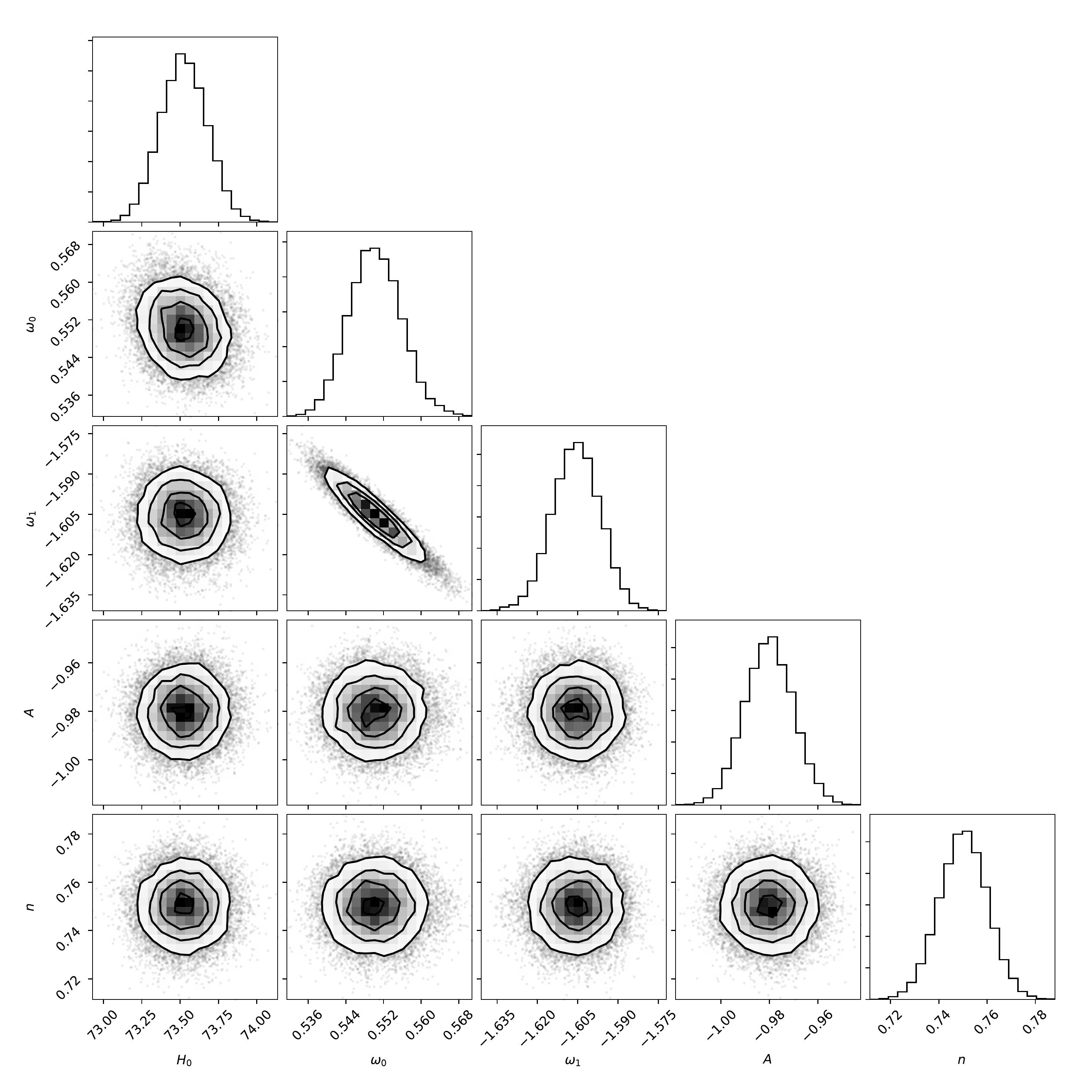}
 \end{array}$
 \end{center}
\caption{Contour map of the model given by Eq.~(\ref{eq:Model2}), for $z \in [0,2.5]$. The best fit values of the model parameters are $H_{0} = 73.52 \pm 0.152$ km/s/Mpc, $ \omega_{0} = 0.55 \pm 0.0055$, $\omega_{1} = -1.605 \pm 0.0087$, $A = -0.98 \pm 0.0098$, and $n = 0.75 \pm 0.01$, when a Bayesian Learning approach based on the generative process has been employed.}
 \label{fig:Fig1_1_a}
\end{figure}

\subsection{Model 2}

Our second model is a modification of Model 1, Eq.~(\ref{eq:Model2}). As will be seen below, it shows that, in addition to the $H_{0}$ tension issue, the problem with higher redshift $H(z)$ data may be solved in an efficient way, too. To construct this second model, the first of its kind, that we know, we have used, in addition, the deceleration parameter, in order to parametrize the inhomogeneity of our fluid. In particular, this second model  has the following form:
\begin{equation}\label{eq:Model2}
P = \left ( \omega_{0} + \frac{\omega_{1}}{1+z} \right ) \rho - A q H^{n},
\end{equation}
where $\omega_{0}$, $\omega_{1}$ and $n$ are the model free parameters, while $q$, $P$ and $\rho$ are the deceleration parameter, the pressure and the energy density of the inhomogeneous fluid, respectively. Again, combining Eq.~(\ref{eq:F1}) with Eq~(\ref{eq:EC}) and taking into account that $\dot{\rho} = - (1+z) H \rho^{\prime}$, after some algebra, we get the following equation
\begin{equation}\label{eq:HP2}
H^{\prime} = \frac{H \left(A (z+1) H^n+3 H^2 (\omega_{0}+\omega_{1}+\omega_{0} z+z+1)\right)}{(z+1)^2 \left(A H^n+2 H^2\right)},
\end{equation}
which describes our model in the generative process. The corresponding contour map, for $z \in [0, 2.5]$, is represented in Fig.~(\ref{fig:Fig1_1_a}), while the contour map for $z \in [0; 5]$ is depicted in Fig.~(\ref{fig:Fig0_1_b}), respectively. A detailed analysis of the model by using a Bayesian Learning approach shows that
\begin{itemize}

\item The best fit values for the model free parameters with $1\sigma$ error is $H_{0} = 73.52 \pm 0.152$ km/s/Mpc, $ \omega_{0} = 0.55 \pm 0.0055$, $\omega_{1} = -1.605 \pm 0.0087$, $A = -0.98 \pm 0.0098$, and $n = 0.75 \pm 0.01$, when $z \in [0,2.5]$.

\item On the other hand, when $z \in [0,5]$, the most likely  best fit values with $1\sigma$ error are: $H_{0} = 73.6 \pm 0.146$ km/s/Mpc, $ \omega_{0} = 0.511 \pm 0.0044$, $\omega_{1} = -1.501\pm 0.0088$, $A = -0.98 \pm 0.01$, and $n = 0.75 \pm 0.01$. 
\end{itemize}

Similarly to the previous case, we have checked,  first of all, the fitting results from the Bayesian Learning approach using available observational $H(z)$ data. Specifically, the best fit values of the model parameters have been used in Eq.~(\ref{eq:HP2}), to get the theoretical expansion rate in order to perform the comparison. The results of this comparison, for the best fit values of the model parameters, are depicted in Fig.~(\ref{fig:Fig0_2})~(right-hand side plot). We clearly see that the model can explain the low redshift $H(z)$ observations very well, and that there is no tension with the higher redshift observations. In other words, the proposed model can explain  the BOSS result, too. On the other hand, the graphical behavior of the deceleration parameter and the equation of state parameter of the fluid, Eq.~(\ref{eq:Model2}), can be seen in Fig.~(\ref{fig:Fig0_3})~(bottom panel). In all cases, only the best fit values of the model parameters have been taken into account.  Moreover, the purple curve  represents the case when $z\in[0,2.5]$, while the dashed red curve, the case when $z\in[0, 5]$. It should be mentioned that the model can explain the late time accelerated expansion and the transition to this phase. Also, we can see that our fluid model, Eq.~(\ref{eq:Model1}), during its evolution will naturally  evolve from a fluid with $P  > 0$ to one with $P < 0$, i.e., an evolving cosmic fluid may cause the emergence of the dark energy responsible for the late time accelerated expansion of our Universe. 

Now, if we take a closer look to the second half of Table~\ref{tab:Table1}, we realize that the $H_{0}$ tension has been solved efficiently. Moreover, we also see that the future measurements of the expansion rate for higher redshift values in $z \in [0,5]$ will, most likely,  significantly affect the parameters $H_{0}$, $\omega_{0}$, and $\omega_{1}$. In particular, we see that the mean and the $1\sigma$ errors for $H_{0}$ and $\omega_{0}$ could be affected considerably; however, only the mean of $\omega_{1}$ might be seriously affected, too. 

In order to continue our discussion, we decided to estimate the mean values of the $P/\rho$ EoS and the deceleration parameter $q$ at $z = 0$. Using the best fit values reported above and summarized in Table~\ref{tab:Table1}, we have found that $P/\rho = -1.057$ and $q = -1.085$, when $z \in [0,2.5]$, while for the whole range $z \in [0,5]$, we encountered that $P/\rho = -0.992$ and $q = -0.989$. Subsequently this means that the transition redshift has been affected, too. Therefore, another interesting difference, to be mentioned,  between Model 1, Eq.~(\ref{eq:Model1}) and Model 2, Eq.~(\ref{eq:Model2}), that will most likely appear when higher redshift $H(z)$ measurements are performed, will be encoded in $\omega(z) = P/ \rho$. Indeed, Model 1, Eq.~(\ref{eq:Model1}), predicts a significant deviation from the cosmological constant value $\omega_{\Lambda} = -1$. However, Model 2, Eq.~(\ref{eq:Model2}), will still most likely  mimic the cosmological constant. However, for the $z\in [0,2.5]$ redshift range, the two models still  provide constraints on the dark energy which are in good agreement with the Planck2018 results. As we mentioned above, the validation of the results obtained for $z\in [0,5]$ from the Bayesian Learning approach are waiting to be validated when new high-redshift $H(z)$ measurements  become available.

\section{Conclusions}\label{sec:conc}

We have here studied two different inhomogeneous, single-fluid models of the Universe  and shown  that the $H_{0}$ tension problem can be effectively solved by using them. Specifically, we have considered the models with $P = \left ( \omega_{0} + \frac{\omega_{1}}{1+z} \right ) \rho - A H^{n}$ and $P = \left ( \omega_{0} + \frac{\omega_{1}}{1+z} \right ) \rho - A q H^{n}$, respectively. Actually, the second model is an extension of the first one, where we take into account that, during the cosmic evolution, the nature of the inhomogeneity can be changed and, in addition, we assume that this can be done by using the deceleration parameter. It is known that the deceleration parameter has changed its sign during the evolution of our Universe. Moreover, it is known that this is not just a simple sign changing process, but that it encodes a very important physical process, which gives birth to dark energy. Latter, we have seen that our approach also softens a hard problem related to the high-redshift behavior of Model 1. Our study is based on a Bayesian Machine Learning approach, which actually does not require real observational data to be used to perform the analysis. The method uses a model based generative process, which allows to constrain the  parameters of the model. In our case, the observable is taken to be the Hubble parameter and making use of the properties of the procedure, we constrain the models for two redshift ranges. Namely, first we have constrained the models for $z\in [0,2.5]$, which covers known $H(z)$ observations. This will be helpful to validate our results in this case. On the other hand, considering the extended redshift range, to be covered by  future collaborations, using the generative process based Bayesian Learning, we have constrained the models for $z\in[0,5]$. The validation of our results for the second redshift range will have to wait a bit, until higher redshift $H(z)$ data are available. Our study shows that inhomogeneous single fluid Universe models can indeed solve the $H_{0}$ tension problem, and that it comes from the mean of the $H_{0}$ parameter values. Indeed, the Bayesian Learning approach puts very tight constraints on the model parameters indicating in this way that the $H_{0}$ tension problem solution is only due to the mean value of $H_{0}$. In this regard, both models have proven to be quite good, as compared with previously considered inhomogeneous fluid models. Our two models can solve the $H_{0}$ tension problem, however, only the one with $P = \left ( \omega_{0} + \frac{\omega_{1}}{1+z} \right ) \rho - A q H^{n}$ is favoured. This is  due to the fact that this model can also explain the BOSS result for the $H(z)$ value at $z=2.34$~(for more discussion about this problem, see Ref.~\cite{MK1} and references therein)\footnote{It should be mentioned that this is due to the best fit values of the model free parameters.}. Another interesting difference between Model 1, Eq.~(\ref{eq:Model1}), and Model 2, Eq.~(\ref{eq:Model2}), is most likely to appear when high redshift $H(z)$ measurements are considered, and it is encoded in $\omega(z) = P/ \rho$. In particular, Model 1, Eq.~(\ref{eq:Model1}), predicts a significant deviation from the cosmological constant case, $\omega_{\Lambda} = -1$. However, Model 2, Eq.~(\ref{eq:Model2}), will most likely still mimic the cosmological constant. Both models, for the $z\in [0,2.5]$ redshift range provide constraints on dark energy in well agreement with the Planck2018 results the higher redshifts in~\cite{Ade3}. Again, the validation of the results obtained for $z\in [0,5]$ from the Bayesian Learning approach will have to  wait until new high-redshift $H(z)$ measurements become available.  

Initial attempts to explore a way to solve the $H_{0}$ tension problem with classical forms of inhomogeneous fluids discussed often in the recent literature have not been very successful. However, the modifications presented above provide a very reasonable solution. In concluding, to be mentioned is that in this paper we have reported a new way to solve the $H_{0}$ tension problem and, at the same time, to make a prediction how the status of the models could change in the near future when new observational data for the expansion rate for $z \in [0,5]$ become available. This has been done using  Bayesian Learning and probabilistic programming, employing a PyMC3 python-based framework. Since the methods used in this study are rather new, we decided to include details on them in an Appendix below. Playing with the form of the fluids considered, we have achieved some remarkable results, we plan to extend to consider more complicated cases. Progress on this will be reported in forthcoming papers.

\section*{Acknowledgements}

This work has been partially supported by MINECO (Spain), project FIS2016-76363-P, and by AGAUR (Catalan Government), project 2017-SGR-247. The work was carried out with the financial support of the Ministry of Education and Science of the Republic of Kazakhstan, Grant No. 0118RK00935 while SDO was visiting ENU.

\section*{APPENDIX}

\begin{figure}[h!]
 \begin{center}$
 \begin{array}{cccc}
\includegraphics[width=120 mm]{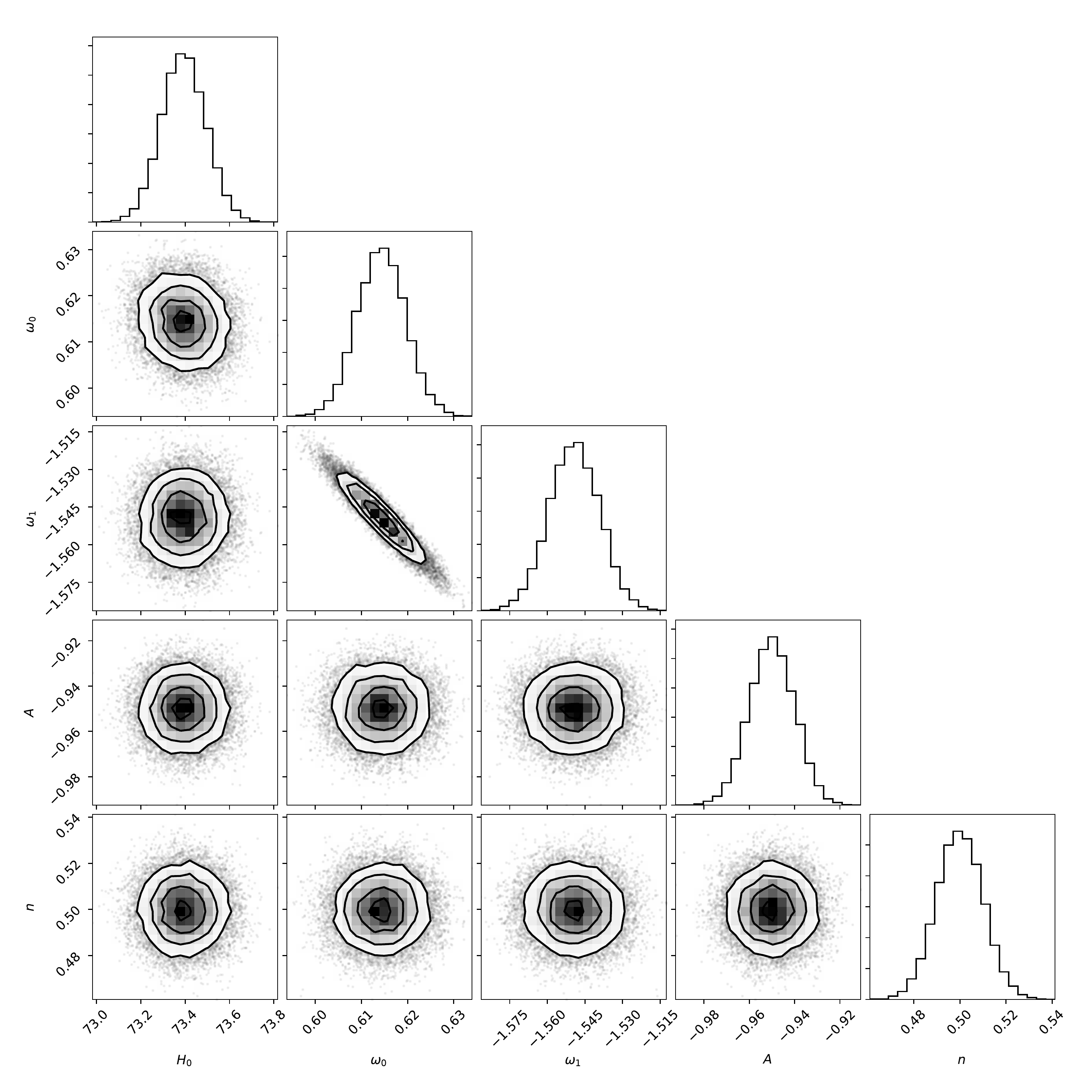}
 \end{array}$
 \end{center}
\caption{Contour map of the model given by Eq.~(\ref{eq:Model1}) for $z \in [0,5]$. The best fit values of the model parameters are $H_{0} = 73.393 \pm 0.1$ km/s/Mpc, $ \omega_{0} = 0.614 \pm 0.0051$, $\omega_{1} = -1.549 \pm 0.01$, $A = -0.95 \pm 0.01$, and $n = 0.49 \pm 0.01$, for $z \in [0,5]$, where a Bayesian Learning approach based on the generative process has been used.}
 \label{fig:Fig0_1_b}
\end{figure}

\begin{figure}[h!]
 \begin{center}$
 \begin{array}{cccc}
\includegraphics[width=120 mm]{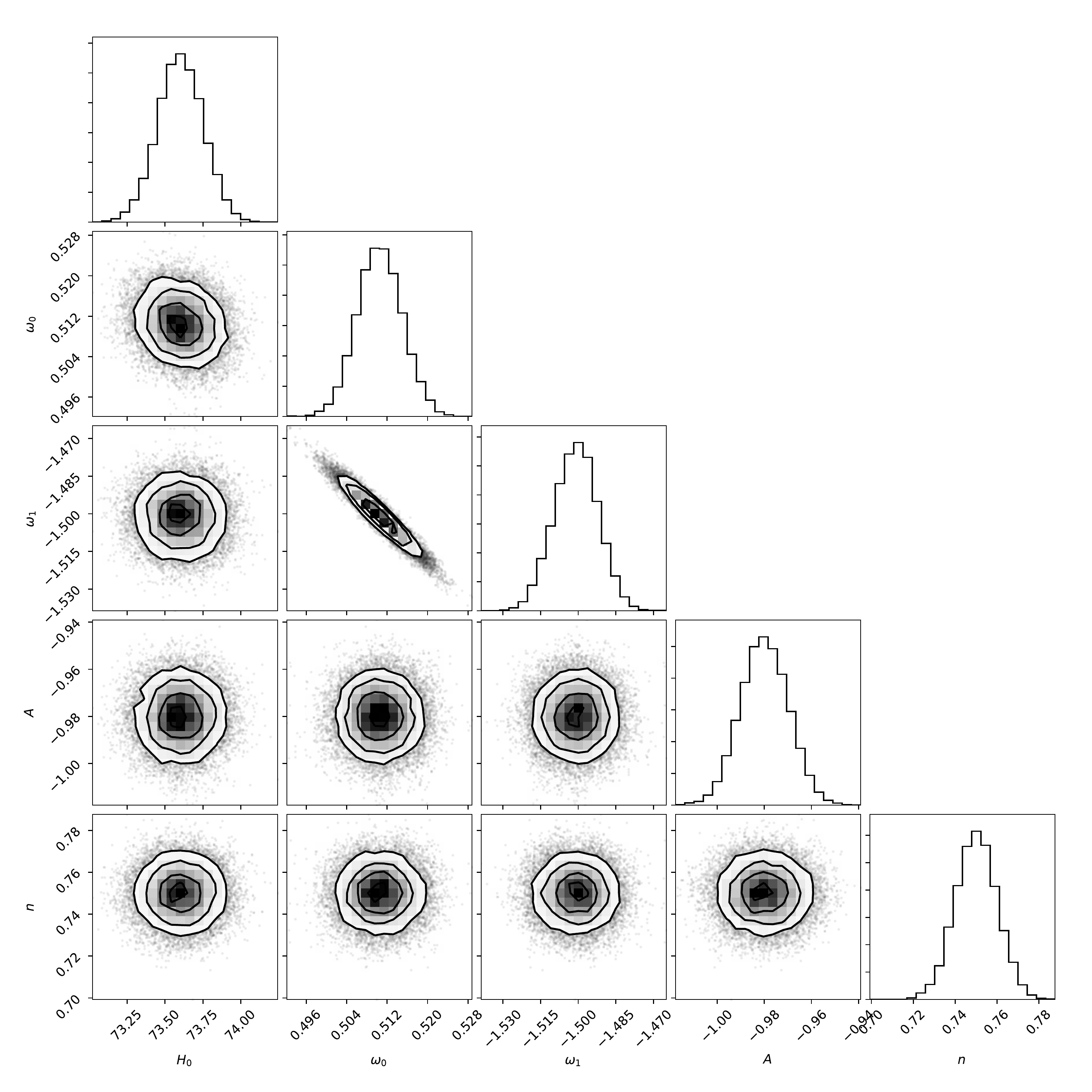}
 \end{array}$
 \end{center}
\caption{Contour map of the model given by Eq.~(\ref{eq:Model2}) for $z \in [0,5]$. The best fit values of the model parameters are $H_{0} = 73.6 \pm 0.146$ km/s/Mpc, $ \omega_{0} = 0.511 \pm 0.0044$, $\omega_{1} = -1.501\pm 0.0088$, $A = -0.98 \pm 0.01$, and $n = 0.75 \pm 0.01$, for $z \in [0,5]$, where a Bayesian Learning approach based on the generative process has been used.}
 \label{fig:Fig1_1_b}
\end{figure}

The contour maps of the Model 1,  Eq.~(\ref{eq:Model1}),  and Model 2, Eq.~(\ref{eq:Model2}), for $z\in [0,5]$ are given  in Figs.~(\ref{fig:Fig0_1_b}) and~(\ref{fig:Fig1_1_b}), respectively. 

\end{document}